\documentclass[aps,preprint,nofootinbib]{revtex4}

\usepackage{graphicx}
\usepackage{epic}
\usepackage{eepic}
\usepackage{latexsym}

\newcommand{\eq}[1]{(\ref{#1})}
\newcommand{\be}{\begin{equation}}
\newcommand{\ee}{\end{equation}}
\newcommand{\bea}{\begin{eqnarray}}
\newcommand{\eea}{\end{eqnarray}}

\newcommand{\vs}[1]{\vspace{#1 mm}}
\newcommand{\hs}[1]{\hspace{#1 mm}}

\def\d{\delta}

\def\f{\phi}
\def\fr{\frac}

\def\l{\lambda}

\def\m{\mu}
\def\n{\nu}

\def\r{\rho}

\def\S{\Sigma}

\def\o{\omega}

\def\del{\partial}
\def\nab{\nabla}

\def\nn{\nonumber}

\newcommand{\dA}{\delta \hs{-.7}A}
\newcommand{\dB}{\delta \hs{-.7}B}
\newcommand{\dC}{\delta \hs{-.7}C}
\newcommand{\dF}{\delta \hs{-.7}\phi}

\begin{document}

\title{Linear Cosmological Perturbations in D-brane Gases}

\author{Ali Kaya}
\email[]{ali.kaya@boun.edu.tr}
\affiliation{Bo\~{g}azi\c{c}i University, Department of
Physics, \\ 34342, Bebek, \.Istanbul, Turkey\vs{3}} 

\date{\today}

\begin{abstract}
We consider linear cosmological perturbations on the background of a 
D-brane gas in which the compact dimensions and the dilaton are
stabilized. We focus on long wavelength fluctuations
and find that there are no instabilities. In particular, the
perturbation of the internal space performs damped oscillations and
decays in time. Therefore, the stabilization mechanism based on D-brane
gases in string theory remains valid in the presence of linearized
inhomogeneities.  
\end{abstract}

\maketitle

\section{introduction}

It is well known that the theory of big-bang nucleosynthesis together
with the observational abundances of light elements provide a limit to
the change in the size of internal space following the production of
these elements. Therefore, string theory should be able to explain why
the extra dimensions are stabilized following nucleosynthesis?
Intuitively, winding branes are expected to resist the expansion and
this may offer an explanation. An accommodation of this idea in early 
string theory can be found in the paper of Brandenberger and Vafa
\cite{bv}, and recently the stabilization of toroidal extra dimensions
by winding strings  is verified by explicit calculations in
\cite{wb,pb} (see also \cite{br}).  The model of
\cite{bv} is developed in \cite{bg3} to include other extended objects
in string theory and this scenario is now known as brane gas cosmology
(BGC) (see e.g. \cite{yeni5}-\cite{t2} for recent work).    

It was also claimed in \cite{bv} that the dimensionality of space-time
can be explained by the annihilation of winding strings in a three
dimensional subspace leading decompactification. From the naive
intersection probability of hypersurfaces embedded in $d$-spatial
dimensions one can infer that $d=2p+1$ is the critical value which
allows annihilation of $p$-branes. Remarkably, for strings
$d=3$ and according to \cite{bv} that is why a three dimensional
subspace is decompactified. As an additional consequence of this
argument, in BGC all higher dimensional branes are expected to be
annihilated in the early universe. Although these claims are supported
numerically in \cite{p1}, there are also some caveats. It is observed
in \cite{n0} that the  number of decompactified dimensions depends on
initial conditions and the outcome of three large dimensions is argued 
to be statistically insignificant \cite{n1}. In \cite{n2} it is shown
that the interaction rates of strings are not strong enough to sustain
thermal equilibrium. Moreover, D-brane anti-D-brane annihilation rate
turns out to be small compared to the expansion rate of a FRW universe
including the higher dimensional D-branes \cite{ann}. Therefore, it
seems that the dimensional counting argument of \cite{bv} is not
sufficient alone for the annihilation process and it is plausible to
have surviving branes other than strings or membranes.   

In the light of these results, we think that late time models with
higher dimensional branes should also be studied in the context of
BGC. Such models may be useful in solving two difficulties with string
gases. Firstly, as pointed out in \cite{yeni4}, strings may not be able
to stabilize volume modulus, or further, there may not be any winding
strings in the spectrum depending on the topology of internal space
(consider e.g. a sphere). Secondly, it turns out to be difficult to
stabilize the dilaton and keep up the stability of extra dimensions
\cite{ds}, which may cause problems in canonical Einstein frame after
compactification.     

In \cite{sa}, we have considered a gas of D6-branes wrapping over the 
six-dimensional internal space and demonstrated dynamical 
stabilization of the volume modulus and the dilaton. The aim of
this paper is to consider long wavelength perturbations in this
model to see the validity of the conclusions under linearized
inhomogeneities. For the string gas the perturbation analysis has been 
carried out in \cite{wb2,sw}. The plan of the paper is as follows. In
the next section we derive linearized equations for the scalar metric
and dilaton perturbations in the generalized longitudinal gauge. 
In section III, we focus on long wavelength fluctuations and
solve the equations. We briefly conclude in section IV.    

\section{Perturbation equations}

We start with ten-dimensional dilaton gravity coupled to matter
Lagrangian ${\cal L}_m$ representing brane winding and momentum modes: 
\be\label{ac}
S=\int\,d^{10}x\,\sqrt{-g}\,\, e^{-2\f}\left[R+4(\nabla\f)^2+e^{a\f}
{\cal L}_m\right]. 
\ee
The field equations that follow from the above action can be found as
\bea
&&R_{\m\n}+2\nabla_\m\nabla_\n\f -\fr{1}{2}g_{\m\n}\left[R+4\nabla^2\f
-4(\nabla\f)^2\right]=e^{a\f}\,T_{\m\n},\label{fe1}\\
&&R+4\nabla^2\f-4(\nabla\f)^2=\fr{(a-2)}{2}\,\,e^\f\,{\cal L}_m,\label{fe2}\\
&&\nabla^\m T_{\m\n} = \fr{(a-2)}{2}\,\,{\cal L}_m \nabla_\n\f-(a-2)T_{\n\l}
\nabla^\l\f,\label{fe3}
\eea
where $T_{\m\n}$ is the matter energy-momentum tensor:
\be\label{lt}
T_{\m\n}=-\fr{1}{\sqrt{-g}}\,\fr{\del}{\del
  g^{\m\n}} \left(\sqrt{-g}\,{\cal L}_m\right).
\ee
Eqns. \eq{fe1} and \eq{fe2}  are obtained by directly varying
the action with respect to the metric and the dilaton. Although for 
matter equations one in principle needs Lagrangian
${\cal L}_m$ explicitly, in our case \eq{fe3} can be deduced from the
consistency of \eq{fe1} and \eq{fe2}.    

It is well known that one particular combination of the field equations
gives a constraint on initial data. Let $n^\m$ to denote the unit normal
vector field to an initial value hypersurface. One can easily
show that \eq{fe1} contracted with $n^\m n^\n$ 
\be
\left(R_{\m\n}+2\nabla_\m\nabla_\n\f -\fr{1}{2}g_{\m\n}\left[R+4\nabla^2\f
-4(\nabla\f)^2\right]-e^{a\f}\,T_{\m\n}\right)n^\m n^\n =0\label{c6}\\
\ee
does not contain any
second order time derivatives. Firstly, it is well from the initial
value formulation of general  relativity that $n^\m
n^\n(R_{\m\n}-g_{\m\n}R/2)$ involves only first order time derivatives
of the metric (see e.g. \cite{wald}). On the other hand, the dangerous
terms with two derivatives acting on dilaton in \eq{c6} can be
reexpressed straightforwardly as    
\bea
\nabla^2\f+n^\m n^\n\nab_\m\nabla_\n\f&=&h^{\m\n}\nab_\m\nab_\n\f\, ,\nn\\
&=&D^\m D_\m\f-(D_\m n^\m)(n^\n\nab_\n\f),\label{6}
\eea
where $g_{\m\n}=h_{\m\n}-n_\m n_\n$ and $D_\m=h_{\m}{}^{\n}\nab_\n$ is
the spatial covariant derivative along constant time hypersurface.
It is clear that \eq{6} contains only first order time derivatives and
thus \eq{c6} can be viewed as a constraint on initial data.   

Assuming a metric of the following form 
\be
ds^2=-e^{2A}dt^2+e^{2B}dx^idx^i+e^{2C}\,d\S_p^2\,\,,
\ee
($i=1,..,m$) the energy-momentum tensor for a gas of branes winding
over the compact manifold $\S_p$ can be found as (see e.g. \cite{ak})
\bea
T_{\hat{0}\hat{0}}&=&T_w\, e^{-mB}\,+\,T_m\,e^{-mB-(p+1)C},\nonumber\\ 
T_{\hat{i}\hat{j}}&=&0,\label{enmom}\\
T_{\hat{a}\hat{b}}&=&-T_w \,e^{-mB}\,\d_{ab}\,+\,\fr{T_m}{p}\,
e^{-mB-(p+1)C}\,\d_{ab},\nonumber 
\eea
where $T_w$ and $T_m$ are constants corresponding to winding and
momentum modes. In this paper we take $\S_p$ to be Ricci flat. From
\eq{enmom} an equation of state $p_{\hat{\m}}=\o_\m\r$ can be deduced where
$\o_i=0,\o_a=-1$ for winding modes and $\o_i=0,\o_a=1/p$ for momentum
modes. Note that the equation for momentum modes is identical to
radiation confined in the compact space. One can check from
\eq{lt} and \eq{enmom} that   
\be
{\cal L}_m=-2\r,
\ee
which is precisely the Lagrangian for hydrodynamical matter (see
e.g. section 10.2 of \cite{rev}). In general we have $\nab^\m
T_{\m\n}\not=0$ due to the coupling of ${\cal L}_m$ to dilaton.

To focus on D-branes one should set 
\be 
a=1,
\ee
which is required  to have the correct $g_s=e^\f$ dependence in the
low energy effective action \eq{ac}. Recall that we also have      
\be
m+p=9,
\ee
where $m$ and $p$ are the dimensions of the observed and the compact
dimensions, respectively. 

In \cite{sa}, we have found a particular solution to \eq{fe1}-\eq{fe3}
\bea
&&ds^2=-dt^2\,+\,(t)^{8/(m+3)}\,dx^idx^i+e^{2C_0}\, d\S_p^2,\nn\\
&&e^\f=e^{\f_0}\,(t)^{2(m-3)/(m+3)}\, ,\label{sol}\\
&&e^{\f_0}=\fr{4\,(p+2)\,p}{T_w\,(p+1)(m+3)^2},\hs{7}
e^{(p+1)C_0}=\fr{(p+2)\,T_m}{p\,T_w},\nn
\eea
which has some appealing
properties. For instance, it is possible to argue
analytically\footnote{As discussed in \cite{t1} and \cite{t2}, it is
  possible to give analogous arguments for intersecting brane
  configurations with or without dilaton.} 
and support the argument numerically that for arbitrary initial
conditions (within a given ansatz) the fields approach \eq{sol}
asymptotically. Moreover, for $m=3$, the dilaton becomes a constant
and thus the canonical Einstein and string frames become identical up 
to a constant scaling. 

Let us now consider linearized inhomogeneities on \eq{sol}. We are
interested in the case where the fluctuations are independent of
compact coordinates\footnote{This is indeed a good approximation since the
extra dimensions are assumed to be very small.} and choose the
generalized longitudinal gauge (see \cite{wb2}) in which the scalar
metric perturbations only appear in diagonal entries.  Therefore,
the perturbed solution has the form: 
\bea
&&ds^2=-e^{2A}(1+2\dA)dt^2\,+\,e^{2B}(1+2\dB)\,dx^idx^i+e^{2C}(1+2\dC)\,
d\S_p^2,\nn\\ 
&&e^\Phi=e^{\f}(1+\dF)\, ,\label{persol}
\eea
where 
$A$, $B$, $C$ and $\f$ denote the background values depending only on
$t$ and 
\be
\dA=\dA(t,x^i),
\hs{5}\dB=\dB(t,x^i),\hs{5}\dC=\dC(t,x^i),\hs{5}\dF=\dF(t,x^i).
\ee 
Let us remind that in the longitudinal gauge $\dA$, $\dB$ and $\dC$
coincide with diffeomorphism invariant quantities. 

In obtaining the linearized field equations we find it convenient to
use \eq{c6} together with a particular combination of \eq{fe1} and
\eq{fe2} 
\be
R_{\m\n}+2\nabla_\m\nabla_\n\f
=\fr{1}{2}g_{\m\n}e^\f\r+e^{\f}\,T_{\m\n}. \label{nfe}
\ee
As we pointed out above, \eq{c6} can be interpreted as a constraint on
initial data .\footnote{It is possible to show that \eq{c6} and
  \eq{nfe} are equivalent to \eq{fe1} and \eq{fe2} with $a=1$.}
Substituting  \eq{enmom} and \eq{persol} in these equations and
keeping only the linear terms one gets 
\bea
&&2\ddot{\dF}-m\, \ddot{\dB}-p\,\ddot{\dC}+e^{-2B}\,\sum_k\dA_{kk}+
2\left[m \ddot{B}+m\, \dot{B}\dot{B}-2\ddot{\f}\right]\dA+\nn\\
&&\left[m\dot{B}-2\dot{\f}\right]
\dot{\dA}-2m\dot{B}\dot{\dB}=\fr12\r e^\f\dF +\fr12
e^\f\d\hs{-.8}\r\, ,\label{e1}\\ 
&&(m-1)\dot{\dB}_i-2\dot{\dF}_i-\left[(m-1)\dot{B}-2\dot{\f}\right]\dA_i
+p\dot{\dC}_i- p\dot{B}\dC_i +2\dot{B}\dF_i=-e^{\f+B}\d T_{0i},
\label{e2}\\
&&\ddot{\dB}+e^{-2B}\left[2\dF_{ii}-\dA_{ii}-p\dC_{ii}-(m-2)\dB_{ii}\right]
-e^{-2B}\sum_k\dB_{kk}+2\left[m\dot{B}-\dot{\f}\right]\dot{\dB}-\nn\\
&&\dot{B}\dot{\dA}+p\dot{B}\dot{\dC}-2\dot{B}\dot{\dF}+2
\left[2\dot{B}\dot{\f}-\ddot{B}-m\dot{B}\dot{B}\right]\dA=\fr12\r e^\f\dF
+\fr12 e^\f\d\hs{-.8}\r\, ,\label{e3}\\
&&\left[2\dF-\dA-p\dC-(m-2)\dB\right]_{ij}=0,\hs{7}i\not=j\label{e4}\\
&&\ddot{\dC}-e^{-2B}\sum_k\dC_{kk}+\left[m\dot{B}-2\dot{\f}\right]\dot{\dC}
=\fr12\r e^\f\dF
+\fr12 e^\f\d\hs{-.8}\r+e^\f T_{aa}\dF+e^\f\d T_{aa}\, ,\label{e5}\\
&&\left[m(m-1)\dot{B}-2m\dot{\f}\right]\dot{\dB}+\left[mp\dot{B}-2p\dot{\f}
\right]\dot{\dC} +
\left[4m\dot{B}\dot{\f}-4\dot{\f}\dot{\f}-m(m-1)\dot{B}\dot{B}\right]
\dA+ \nn\\
&&\left[4\dot{\f}-2m\dot{B}\right]\dot{\dF}+e^{-2B}\sum_k
\left[2\dF_{kk}-(m-1)\dB_{kk}-p\dC_{kk}\right]
=\r e^\f\dF +e^\f\d\hs{-.8}\r\, ,\label{incon}
\eea
where \eq{e1}-\eq{e5} follow from \eq{nfe}, and \eq{incon} can be
obtained from \eq{c6}. In the above equations the dot and the
subindex $i$ denote partial differentiations with respect to $t$ and
$x^i$, respectively, and we have not used the summation convention.

Note that in \eq{e2} we introduce a non-diagonal component for the
linearized energy momentum tensor $\d T_{0i}$. Although for
long wavelength fluctuations this equation is satisfied identically, it
is in general necessary to  add $\d T_{0i}$ since the
perturbed metric functions depend on both $t$ and $x^i$, which gives 
non-zero $R_{0i}$ (indeed \eq{e2} is the $(0i)$ component of
\eq{nfe}). Moreover, one can see that without $\d T_{0i}$ the matter
equations constrain the metric functions too much, see \eq{lc2}
below. The appearance of this non-diagonal term for winding modes
can be deduced from the Born-Infeld (BI) action as follows. When the
metric functions depend only on $t$, a static brane winding
$\S_p$ and located at a constant position $x^i=x^i_0$ becomes an
extremum of the BI action (i.e. the embedding coordinates become
harmonic maps). Using this solution in the energy-momentum tensor
obtained from BI action and smearing the delta function singularity by
taking a continuum average for a gas of branes, 
one gets the winding part of \eq{enmom}
\cite{bg12}. However, when the metric is perturbed with $t$ and $x^i$
dependent functions, the static brane solution should
be modified $x^i=x^i_0+\d x^i(t)$ and this correction induces
non-zero $\d T_{0i}$.     

Linearizing the matter equation \eq{fe3} we find that $\n=a$ component
is identically satisfied while $\n=0$ and $\n=i$ equations give  
\bea
&&\dot{\d\r}+(m\dot{B}+p\dot{C})\d\r+(m\dot{\dB}+p\dot{\dC})\r+
\dot{\dC\,}T^{a}{}_{a}+\dot{C}\,\d T^{a}{}_{a}+e^{A-B}
\d T^{0j}_{,j}=0,\label{lc1}\\
&&\dot{\d T_{0i}}+\left[(m+1)\dot{B}+p\dot{C}-\dot{\f}\right]\d T_{0i}
+e^{A-B}\left[\dC_i T^{a}{}_{a}+\r\dF_i-\r\dA_i\right]=0.
\label{lc2}
\eea
Eq. \eq{lc1} has actually two separate parts corresponding to winding
and momentum modes. In obtaining $\d\r$ and $\d T_{ab}$ from
\eq{enmom}, one should also consider the variations $\d T_w$ and $\d
T_m$. Therefore, there are seven variables which are $\d T_w$, $\d
T_m$, $\d T_{0i}$, $\dA$, $\dB$, $\dC$ and $\dF$. Equations \eq{lc1} and
\eq{lc2} determine the evolution of source fields $\d T_w$, $\d
T_m$ and $\d T_{0i}$, which in turn should
be substituted in \eq{e1}-\eq{incon}. It is possible to
solve \eq{e4} as 
\be
2\dF-\dA-p\,\dC-(m-2)\dB=0,\label{coz}
\ee
which reduces the number of unknown functions by one. On the other
hand, \eq{e2} and \eq{incon} are not dynamical evolution equations 
since they only contain first order time derivatives and thus merely
restrict the solution space. Ignoring them, it is easy to see that the
number of equations and unknowns are the same.      

\section{Long wavelength fluctuations}

Let us now focus on long wavelength perturbations. As pointed out in
\cite{wb2} an instability on such scales would be a problem for the
stabilization mechanism. In this regime all spatial derivatives can be
ignored in field equations which in turn implies that \eq{e2} and
\eq{lc2} are identically satisfied. On the other hand, from \eq{lc1}
one obtains  
\be\label{zero}
\dot{\d T_w}=\dot{\d T_m}=0.
\ee
Since constant shifts of $T_w$ or $T_m$ alter the  solution \eq{sol},
they are not honest perturbations and one should set $\d T_w=\d
T_m=0$. In other words, \eq{zero} gives the zero modes which must be
ignored. Solving $\dF$ from \eq{coz}\footnote{For long wavelength
  fluctuations \eq{coz} can be viewed as the residual gauge
  symmetry fixing.} and using the background values from \eq{sol}, we find
that \eq{e1},  \eq{e3}, \eq{e5} and \eq{incon} respectively become    
\bea
&&\ddot{\dA}-2\ddot{\dB}+\fr{12\dot{\dA}}{(m+3)t}-\fr{8m\dot{\dB}}{(m+3)t}
-\fr{10p\dA}{(m+3)^2t^2}+\fr{2p(m+2)\dB}{(m+3)^2t^2}=0,\label{l1}\\
&&\ddot{\dB}-\fr{8\dot{\dA}}{(m+3)t}+\fr{20\dot{\dB}}{(m+3)t}
-\fr{10p\dA}{(m+3)^2t^2}+\fr{2p(m+2)\dB}{(m+3)^2t^2}=0,\label{l2}\\
&&\ddot{\dC}+\fr{12\dot{\dC}}{(m+3)t}+\fr{2p(p+2)\dC}{(m+3)^2t^2}=0,
\label{l3}\\
&&\fr{12\dot{\dA}}{(m+3)t}+\fr{4(m-6)\dot{\dB}}{(m+3)t}
+\fr{20p\dA}{(m+3)^2t^2}-\fr{4p(m+2)\dB}{(m+3)^2t^2}=0.\label{l4}
\eea
Note that $\dC$ decouples from $\dA$ and $\dB$. As a consistency
check of these equations, we find that the time derivative of the
constraint \eq{l4} is proportional to four times \eq{l1} plus $m$
times \eq{l2} and thus identically satisfied, as it should be. This
does not mean however that \eq{l4} is empty. Rather, one should  first
solve \eq{l1}-\eq{l2} and then use \eq{l4} to narrow the solution
space.       

Eq. \eq{l3} can easily be solved for $\dC$ which gives 
\be
\dC(t)=c_1\,\fr{\cos(a\ln t)}{t^b}+c_2\,\fr{\sin(a\ln t)}{t^b},
\ee
where $a,b$ are positive numbers depending on $m$ (e.g. for $m=3$,
$a=\sqrt{29/12}$, $b=1/2$). On the other hand, the coupled equations
\eq{l1} and \eq{l2} give  
\bea
&& \dA=\,(m+2)\,k_1+k_2\,
t^{-(7+m)/(m+3)}+k_3\,(m-2)t^{(-9+m)/(m+3)}+k_4\,(29+11m)\,t^2\nn\\
&&\dB=5k_1-k_2\,t^{-(7+m)/(m+3)}+k_3\,t^{(-9+m)/(m+3)}+k_4(23+m)\,t^2
\label{ms}
\eea
where $k_1,..,k_4$ are arbitrary constants. Substituting these solutions
into \eq{l4}, we observe that it yields 
\be
k_4=0,
\ee
and thus the growing mode in \eq{ms} is eliminated by the constraint
equation. As a result, we find that there are no instabilities to this
order. 

\section{Conclusions}

In this paper, we consider long wavelength fluctuations on the
background of a D-brane gas in ten dimensional dilaton gravity
constructed in \cite{sa}. The most attractive feature of the
solution is that it yields stabilized internal dimensions. Moreover,
when the observed space is three dimensional the dilaton also becomes a
constant. The main result of this paper is that the long wavelength
perturbation of the internal space performs damped oscillations. Other
fluctuations, including the dilaton, are found to decay to a constant.
Therefore, the stabilization mechanism based on D-brane winding
and momentum modes remains valid in the presence of linearized
inhomogeneities.  

Technically, the perturbation analysis for D-branes is similar to the string
gases studied in \cite{wb2}. The main difference is that in our case
there appears to be a growing mode which is eliminated by the
constraint equation. Although the presence of this  mode would not
affect the internal space, it would ruin the stabilization of dilaton
for $m=3$, which is crucial in the effective field theory obtained
after compactification. 

Higher dimensional D-brane gases offer an analogous scenario to string
gases for the stabilization of the extra dimensions. Moreover, D6
branes happen to fix dilaton in three dimensions. 
This mechanism can be useful in
topologically non-trivial  compactifications where the winding strings
do not appear in the spectrum or they are not capable of fixing volume
modulus. Of course, the toy model we consider is immature and it
should be developed in many different directions. However, the results
in BGC show that string theory has the right ingredients to solve the
stabilization problem of extra dimensions in cosmology.

\end{document}